\documentclass[11pt]{article}
\usepackage[pctex32]{graphics}
\usepackage{amsmath}
\usepackage{amssymb,latexsym}
\usepackage{setspace}
\usepackage[mathscr]{eucal}

\newcommand{\bee}{\begin{equation}}
\newcommand{\ene}{\end{equation}}
\newcommand{\pa}{\partial}
\newcommand{\al}{\alpha}

\begin{document}

\vspace*{1in}

\begin{center} {\large\bf Vortex Stretching and Reconnection in a Compressible Fluid} \\

\vspace{1in}

Bhimsen K. Shivamoggi\\
University of Central Florida\\
 Orlando, FL 32816

\vspace{.5in}

\end{center}

{\bf Abstract}

Vortex stretching in a  compressible fluid is considered. Two-dimensional (2D) and  axisymmetric cases are considered   separately. The flows associated with the vortices are   {\it perpendicular} to the plane of the uniform   straining flows. Externally-imposed density build-up near the axis leads to enhanced compactness of the vortices -- ``dressed" vortices (in analogy to ``dressed" charged particles in a dielectric system). The compressible vortex flow solutions in the 2D as well as axisymmetric cases identify a length scale relevant for the compressible case which leads to the Kadomtsev-Petviashvili spectrum for compressible turbulence. Vortex reconnection process in a compressible fluid is shown to be possible even in the inviscid case -- compressibility leads to defreezing of vortex lines in the fluid.


\section{Introduction}

The vortex stretching process - 
\begin{itemize}
\item leads to the transport of energy among   various scales of motion in a turbulent flow,
\item plays an important role in the vortex   reconnection process   and hence in describing the fine scales of turbulence.
\end{itemize}

Vortex reconnection (Siggia and Pumir \cite{sp85}, Schatzle \cite{prs87}) has been argued to be a prime candidate for a {\it finite-time}  singularity in Euler equations. Such a singularity plays a central role in the small-scale dynamics of turbulence by   producing arbitrarily large velocity gradients. However, vortex reconnection is a process that is not yet well understood.  Certain ``canonical" cases of vortex reconnection have been investigated in great detail,   both  experimentally (Fohl and Turner, \cite{ft75},  Oshima and Asaka \cite{oa77}) and  numerically   (Ashurst and Meiron \cite{am87}, Pumir and Kerr, \cite{pk87}, Kida and Takaoka \cite{kt88} and others). But, a {\it global} view of the various  reconnection scenarios is not at hand yet.
  
Laboratory Experiments (Cadot et al. \cite{cdc95}, Villermaux et at. \cite{vsg95}) and DNS (Jimenez et al. \cite{jwsr93}) have revealed strong  coherent and elongated vortices among the small scales in incompressible turbulence. These vortices are believed to originate from strained vorticity fields like the {\it Burgers} vortex (Burgers \cite{jmb51}). 
Burgers vortex describes the interplay between  the intensification of vorticity due to the imposed straining flow and the diffusion of vorticity due to the action of viscosity. The straining simulates {\it locally} the stretching undergone by each vortex in the velocity field induced by other vortices. Intermittency structures that exhibit velocity profiles similar to that of Burgers vortex have been observed in  grid turbulence (Mouri et al.  \cite{mhk00}).

The two-dimensional (2D) Burgers vortex solution is of the form (Robinson and Saffman \cite{rs84}) - 
\bee
{\bf v}={\left\{-\alpha x+u(x,y,t),-\beta y+v(x,y,t),(\alpha+\beta)z\right\}}.
\ene 
The quantity $(\alpha+\beta)$  $(\alpha$ and $\beta>0)$
measures the stretching rate of vortices, which are aligned along the $z$-axis (that is also the principal axis of a uniform plane straining flow).
 Numerical solutions of three-dimensional (3D) Navier-Stokes equations (Ashurst et al. \cite{akkg87} and others) have confirmed the alignment between the  vorticity and one principal axis of the local strain. The velocity induced by the vorticity lies in the $xy$-plane, with components $u$ and $v$ which are independent of $z$. Simple closed-form steady solutions exist for the following special cases - 
\vspace{-.1in}
\begin{itemize}
\item $\al=\beta>0$ - axisymmetric vortex
\vspace{-.2in}
\item $\al>0,\beta= 0$ - 2D shear layer.
\end{itemize}
Robinson and Saffman \cite{rs84} demonstrated the existence of solutions for arbitrary values of the ratio $\al /\beta$.

Unsteady 2D Burgers vortex solutions have   been used to model the spanwise structure of turbulent mixing layers (Lin and Corcos \cite{lc84}, Neu \cite{jn84}). Unsteady axisymmetric Burgers vortex solutions have been used to model the fine-scale structure of homogeneous incompressible turbulence (Townsend \cite{aat51}, Lundgren \cite{tsl82}).
 
DNS (Porter et al. \cite{ppw98}) have confirmed the existence of vortex filaments in compressible turbulence. The vortex stretching process can be expected to be influenced in an essential way by fluid compressibility (Shivamoggi \cite{bks99} and \cite{bks02}). So, investigation of stretched vortices in a compressible fluid is in order which is addressed in this paper along with applications to compressible turbulence.

Vortex reconnection in a compressible fluid is a topic in its infancy (Virk et al. \cite{vhk95} and Shivamoggi \cite{bks96}). Additional mechanisms of vorticity generation like the {\it baroclinic} vorticity generation exist in a compressible fluid. The vortex reconnection process in a compressible fluid is therefore more complicated than its counterpart in an incompressible fluid. Further exploration of the basic mechanism underlying this process is in order and is addressed in general terms in this paper.

\section{Modified 2D Burgers Vortex}

Consider a modified Burgers vortex flow with the velocity field given by
\bee
{\bf v} = \{-\gamma(t)x,\gamma(t)y,W(x,t)\}.
\ene
(2) describes the convection of the vortex lines toward the $y$-axis and the stretching along the $y$-axis by the imposed straining flow. The straining flow is externally imposed, so the vorticity is decoupled from the dynamics of the  straining flow that is stretching it. The streamlines (see Figure 1) shown in the $x,y$-plane   represent the uniform plane straining flow. This streamline pattern is the same in each plane parallel to the $ x,y$-plane. Observe that the flow associated with the vortex in question is {\it perpendicular} to the plane of the uniform  straining flow, unlike the Burgers vortex given by (1). This situation is well-suited for modelling a mixing-layer flow or jet flow.  (2) describes the convection of the vortex lines towards the $x=0$ plane and the stretching in the $y$-direction by the imposed straining flow. \\

\begin{center}
{\includegraphics{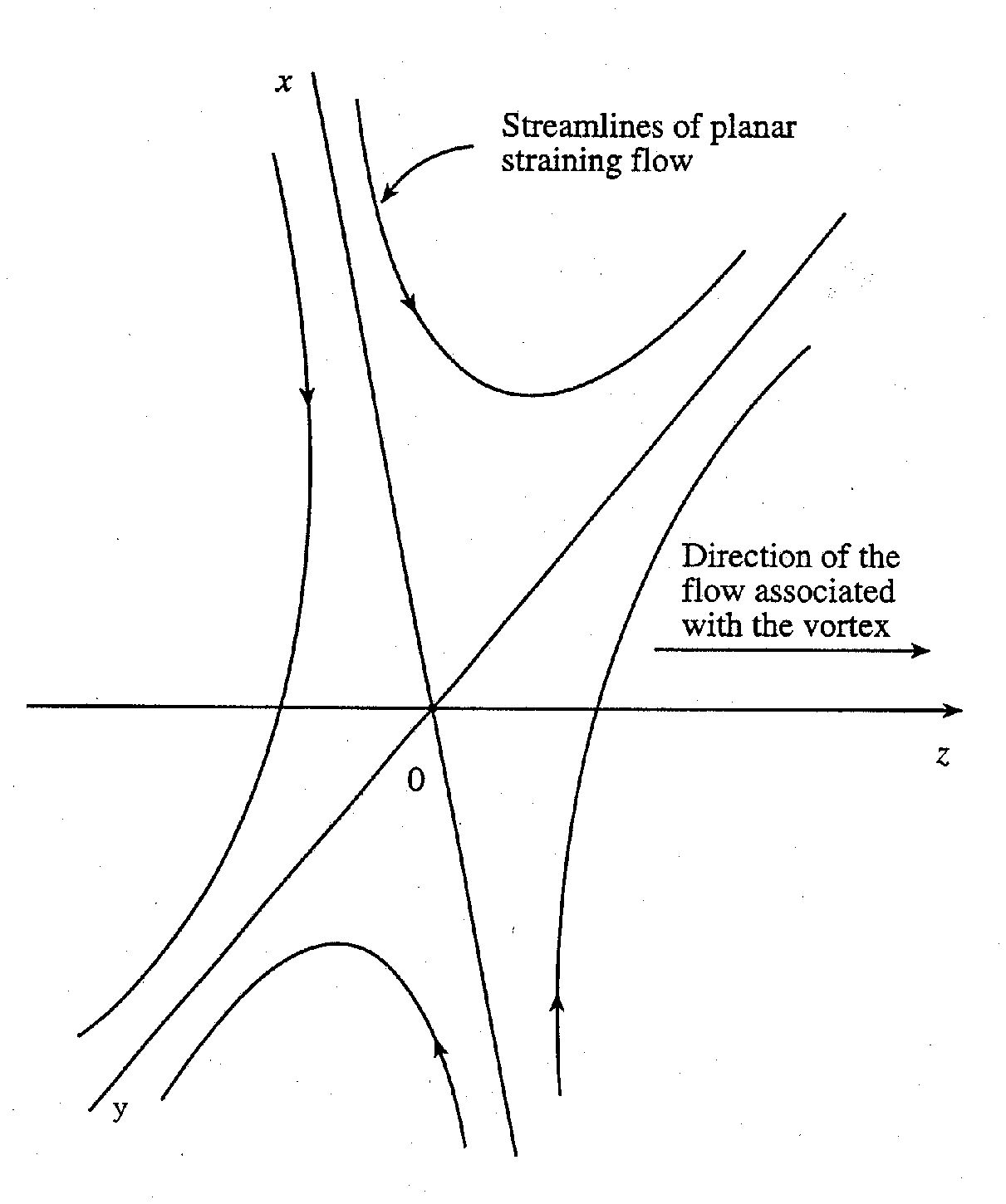}}
\end{center}

\begin{center} Figure 1. Modified Burgers Vortex Model. \hspace{.5in}
\end{center}


The vorticity field corresponding to (2) is
\bee
{\boldsymbol \omega} = \nabla\times{\bf v}=\left\{0,-\frac{\pa W}{\pa x},0\right\}
\ene
which shows that the vortex lines for this model are aligned along the $y$-axis which happens to be the principal axis of a uniform plane  straining flow (2), however,  as in the Burgers   vortex model (1).

Using (2) and (3), the vorticity conservation equation
\bee
\frac{\pa{\boldsymbol \omega}}{\pa t}+({\bf v} \cdot \nabla ){\boldsymbol \omega}=(\boldsymbol \omega \cdot \nabla ){\bf v} +\nu \nabla^2 {\boldsymbol \omega}
\ene
becomes
\bee
\frac{\pa\Omega}{\pa t}-\gamma x\frac{\pa\Omega}{\pa x} =\gamma\Omega +\nu \frac{\pa^2\Omega}{\pa x^2}
\ene
where $\nu$ is the kinematic viscosity and $\Omega$ is the vorticity-
\bee
\Omega\equiv\frac{\pa W}{\pa x} \, .
\ene

Introducing dimensionless independent variables -
\bee
\xi =\sqrt{\frac{\gamma }{\nu}} \; x, \quad \tau =\int^t \gamma (t')dt'
\ene
equation (5) becomes
\bee
\frac{\pa}{\pa\xi} \left(\frac{\pa \Omega}{\pa\xi} +\xi\Omega\right)=\frac{\pa\Omega}{\pa\tau} \, .
\ene

Let the boundary conditions be 
\bee
|\xi | \Rightarrow\infty :\Omega\Rightarrow 0.
\ene


\noindent
{\bf (i) Steady Case:}

For the steady case (with $\gamma$ = constant), equation (8) becomes
\bee
\frac{d}{d\xi} \left(\frac{d\Omega}{d\xi} +\xi\Omega\right)=0.
\ene

Using (9), equation (10) has the solution -
\bee
\Omega =c_1 e^{-\xi^2/2}
\ene
or
\bee
W(\xi )=c_1 erf (\xi / \sqrt{2})
\ene
which represents the shear layer.

For this shear-layer flow solution, the build-up   of vorticity due to the  convection   of the vortex lines towards the  $x=0$ plane and the stretching in the $y$-direction by the imposed straining flow is counterbalanced by the diffusion of vorticity in the $x$-direction.
 
\medskip
\noindent
{\bf (ii) Unsteady Case:}

For the unsteady case, let us look for a solution of the form -
\bee
\Omega (\xi ,\tau )=h_\lambda (\xi )e^{-\lambda\tau} .
\ene
Equation (8) then yields
\bee
\frac{d}{d\xi} \left(\frac{dh_\lambda}{d\xi} +\xi h_\lambda \right) =-\lambda h_\lambda .
\ene

For bounded solutions of equation (14) to   exist, we require
\bee
\lambda =n; \quad n=0,1,2,\dots \; .
\ene
Equation (14) then has the solution -
\bee
h_n(\xi )=(-1)^n h_0(\xi ) H_n (\xi ); \quad n=0,1,2,\dots
\ene
where,
$$
h_0(\xi )=e^{-\xi^2 /2}
$$
and $H_n(\xi)$ are the Hermite polynomials -
$$
H_0(\xi) = 1,\;\; H_1(\xi )= \xi ,\;\; H_2(\xi )= \xi^2-1,\;\; H_3(\xi )= \xi^3-3\xi,\;\; \text{etc.}
$$

Observe that $n=0$ (steady case) corresponds to the shear-layer solution (11) while $n=1$ (unsteady case) corresponds to the jet solution.

\section{Compressible Modified 2D Burgers   Vortex} 

Let us now consider the modified 2D Burgers vortex in a compressible barotropic fluid. For this purpose, let the velocity and density profiles be given by (Shivamoggi \cite{bks99}) -
\begin{subequations}
\begin{align}
{\bf v} &= \{ \dot \al (t)x, \dot \beta (t)y,W(x,t)\}\\
\rho &=\sigma (t)+\frac{\rho_0}{U} (\dot \al +\dot \beta )x .
\end{align}
\end{subequations}
where $\rho_0$ and $U$ are reference density and   velocity, respectively. 
(17) describes a density build-up (or decay) in the direction along which  vortex lines are being compressed by the imposed straining flow. This arrangement  maximizes compressibility  effects on the vortex stretching process.
 
Using (17), the mass-conservation equation
\bee
\frac{\pa\rho}{\pa t}+({\bf v}\cdot\nabla )\rho +\rho (\nabla \cdot{\bf v})=0
\ene
yields
\bee
\dot \sigma +\frac{\rho_0}{U} (\ddot \al + \ddot \beta )x+\dot \al \frac{\rho_0}{U} x(\ddot \al +\ddot \beta )  +[\sigma +\frac{\rho_0}{U} (\dot \al + \dot \beta )x] (\dot \al +\dot \beta )=0
\ene
from which, we obtain the following relations - 
\begin{gather}
\dot \sigma +\sigma (\dot \al +\dot \beta )=0\\
(\ddot \al +\ddot \beta )+\dot \al (\dot \al +\dot \beta )+(\dot \al +\dot \beta )^2 =0.
\end{gather}

Using equation (20), equation (21) becomes
\bee
\frac{d}{dt} \left(\frac{\dot\sigma}{\sigma}\right) +\dot\al \left(\frac{\dot\sigma}{\sigma}\right) -\left(\frac{\dot\sigma}{\sigma}\right)^2=0.
\ene

Next, using (17),  the vorticity conservation   equation
\bee
\rho \left[\frac{\pa{\boldsymbol \omega}}{\pa t} +({\bf v} \cdot\nabla ){\boldsymbol \omega} \right] +\nabla\rho\times \left[ \frac{\pa{\bf v}}{\pa t} +({\bf v}\cdot\nabla ){\bf v} \right]  =\rho ({\boldsymbol \omega} \cdot\nabla ){\bf v} -\rho (\nabla\cdot {\bf v} ){\boldsymbol \omega} +\mu \nabla^2 {\boldsymbol \omega}
\ene
leads to 
\begin{align}
&\left[ \sigma +\frac{\rho_0}{U} (\dot \al +\dot \beta )x\right] \left[ \frac{\pa^2 W}{\pa x\pa t} +\dot \al x\frac{\pa^2W}{\pa x^2}\right]  +\frac{\rho_0}{U} (\dot\al +\dot\beta ) \left[ \frac{\pa W}{\pa t}+\dot\al x\frac{\pa W}{\pa x}\right]\notag\\
&\hspace{2.3in} =-\dot\al \frac{\pa W}{\pa x} \left[\sigma +\frac{\rho_0}{U} (\dot\al +\dot\beta )x\right] +\mu \frac{\pa^3 W}{\pa x^3}
\end{align}
\bee
(\dot\al +\dot\beta )(\ddot\beta +\dot\beta^2 )=0
\ene
where $\mu$ is the dynamic viscosity.

Using (20), and the $z$-component of the  
{\it equation of motion}, namely,
\bee
\rho\left(\frac{\pa W}{\pa t} +\dot\al x\frac{\pa W}{\pa x}\right)=\mu \frac{\pa^2 W}{\pa x^2}
\ene
equations (24) and (25) become
\bee
\left[ \sigma -\frac{\rho_0}{U}\left(\frac{\dot\sigma}{\sigma}\right)x\right] \left[\frac{\pa\Omega}{\pa t}+\dot\al x\frac{\pa \Omega}{\pa x} +\dot\al\Omega\right] -\frac{
\frac{\rho_0}{U}
\left(\frac{\dot\sigma}{\sigma}\right)}
{\left[ \sigma -\frac{\rho_0}{U} 
\left(\frac{\dot\sigma}{\sigma} \right) x
\right]} \mu \frac{\pa\Omega}{\pa x} = \mu\frac{\pa^2\Omega}{\pa x^2}
\ene
\bee
\frac{\dot\sigma}{\sigma} (\ddot\beta +\dot\beta^2 )=0.
\ene

For the compressible case, $\dot\sigma /\sigma \neq 0$, so   equation (28) reduces to 
\bee
\ddot\beta +\dot \beta^2=0.
\ene

In order to facilitate an analytic solution,   consider the case
\bee
\sigma =e^{\int^t c(t')dt'}, \quad \dot\al =-a, \quad \dot\beta =b.
\ene

Equations (20), (22) and (29) then yield a closed set of equations for $a(t), b(t)$ and $ c(t)$ (which are the same as those for the 2D case):
\begin{align}
&b(t)=\frac{1}{t+A},\quad c(t)=\frac{e^{\int^t a(t')dt'}}{B-\int^t e^{\int^{t'} a(t^{\prime\prime} )dt^{\prime\prime}} dt'} ,\notag\\
& a(t)=c(t)+b(t),
\end{align}
where $A$ and $B$ are arbitrary constants.

Equation (27) then becomes
\bee
\left(\rho_0 e^{\int^t c(t')dt'} -\frac{\rho_0}{U} cx\right) \left(\frac{\pa\Omega}{\pa t}-ax\frac{\pa\Omega}{\pa x} -a\Omega\right) -\frac{\left(\frac{\rho_0}{U} \right)c}{\left(\rho_0 e^{\int^t c(t')dt'} -\frac{\rho_0}{U} cx\right)} \mu\frac{\pa\Omega}{\pa x} =\mu \frac{\pa^2\Omega}{\pa x^2} .
\ene

In order to simplify equation (32), let us   assume that the  compressibility effects are weak. From (20) and (31), this implies that $c(t)$, which is a measure of the density change, is small. Equation (22) then leads to
\bee
\frac{dc}{dt} -a(t)c\approx 0
\ene
from which, 
\bee
c(t)\approx c_0e^{\int^t a(t')dt'}
\ene
$c_0$ being an arbitrary constant. (34) replaces the second of the three solutions in (31). Further, in the weakly-compressible case, equation (28) (which, to first approximation, is   automatically satisfied) does not lead to   equation (29) which, therefore, has to be   abandoned. This implies that the first of the three solutions in (31), which comes from equation (29), has to be dropped, as in the 2D case.

Thus, keeping only terms of $O(c)$, and introducing dimensionless independent variables -
\bee
\xi\equiv \sqrt{\frac{a}{\mu}} \; x, \quad \tau\approx\int^t a(t')dt'
\ene
equation (32) may be approximated by 
\bee
\frac{\pa}{\pa\xi} \left[ \frac{\pa\Omega}{\pa\xi} +\xi\Omega \right] =\frac{\pa\Omega}{\pa\tau} - \tilde c \frac{\pa\Omega}{\pa\xi}
\ene
where,
$$
\tilde c(\tau )\equiv \frac{c(t(\tau ))}{\rho_0 U} \sqrt{\frac{\mu}{a(\tau )}} \; .
$$
The boundary conditions on $\Omega$ are the same as in (9).

Comparison of equation (36) with the   corresponding equation (8) for the incompressible case shows that the last term on the right hand side in equation (36) represents the {\it first}-order contribution due to the  compressibility effects (assumed to be weak). Further, observe that the compressibility   effects impart hyperbolic character to equation (36), associated with sound-wave propagation in the fluid.

As a first approximation, if we ignore the time-dependence of the straining-flow profiles, and hence, $\tilde c(\tau)$, and treat $\tilde c (\tau)$ as  a constant,   equation (36) can be solved exactly to give
\bee
\Omega (\tau ,\xi )\approx (-1)^n e^{-\frac{(\xi +\tilde c )^2}{2}} H_n (\xi +\tilde c )e^{-n\tau}
\ene
where $H_n(\xi )$ are the {\it Hermite} polynomials.

Comparison of the compressible vortex profile (37) with the corresponding compressible   vortex profile (13), (15) and (16) shows that, for the 2D case, the {\it first}-order effect of compressibility is to cause a mere {\it Galilean} translation in space of   the  incompressible vortex profiles. Therefore, in order to capture non-trivial effects of compressibility in the 2D case one needs to consider the  time dependence of the straining-flow   profiles. This restriction turns out to be relaxed for the axisymmetric case (below).

\section{Compressible Axisymmetric Stretched Vortex}

Consider an axisymmetric stretched vortex in a compressible barotropic fluid (Shivamoggi \cite{bks02}). Let the velocity (in cylindrical polar   coordinates $(r,\theta ,z)$)  and the density profiles be given by\footnote{A general class of velocity-field profiles of which (38a) is a special case has been discussed by Ohkitani and Gibbon \cite{og00}.} -
\begin{subequations}
\begin{align}
{\bf v} &= \{ \dot\al (t)r,W(r,t),\dot\beta (t)z\}\\
\rho &= \sigma (t)+\frac{\rho_0}{U} (2\dot\al +\dot\beta )r
\end{align}
\end{subequations}
where $\rho_0$ and $U$ are reference density and   velocity, respectively. (38) describes a density build-up (or decay) towards the axis (which is also the direction along which vortex lines are being compressed by the imposed straining flow). 

The vorticity field corresponding to (38) is
\bee
{\boldsymbol \omega} =\nabla\times {\bf v} =\{ 0,0,\Omega\} ,
\ene
where
$$
\Omega =D_rW, \quad D_r \equiv \frac{\pa}{\pa r}+\frac{1}{r} \, .
$$

(39) shows that the vortex lines for this model are aligned along the $z$-axis, which happens to be the direction of the principal extensional strain associated with the axisymmetric uniform straining flow (38).  Further, the flow associated with the vortex is again perpendicular to the plane of the uniform straining flow, a situation that is well suited to modeling an axisymmetric mixing-layer flow. (38) describes the  convection of the vortex lines towards the $z$-axis and the  stretching along the $z$-axis by the imposed straining flow.

Using (38), the mass-conservation equation (18) yields
\bee
\dot\sigma +\frac{\rho_0}{U} (2\ddot\al +\ddot\beta )r+\dot a r\frac{\rho_0}{U} (2\dot\al +\dot\beta ) +\left[ \sigma +\frac{\rho_0}{U} (2\dot\al +\dot\beta )r\right] (2\dot\al +\dot\beta )=0
\ene
from which we obtain the following relations
\begin{gather}
\dot\sigma +\sigma (2\dot\al +\dot\beta )=0\\
(2\ddot\al +\ddot\beta )+\dot\al (2\dot\al +\dot\beta )+(2\dot\al +\dot\beta )^2=0.
\end{gather}

Using equation (41), equation (42) becomes
\bee
\frac{d}{dt}\left(\frac{\dot\sigma}{\sigma}\right) +\dot\al \left(\frac{\dot\sigma}{\sigma}\right) -\left(\frac{\dot\sigma}{\sigma} \right)^2 =0.
\ene

Next, using (38), the vorticity-conservation   equation (23) leads to
\begin{align}
&\left[ \sigma +\frac{\rho_0}{U} \left( 2\dot\al +\dot\beta\right) r\right] \left[ \frac{\pa\Omega}{\pa t} +\dot\al r\frac{\pa\Omega}{\pa r} \right]  + \frac{\rho_0}{U} \left( 2\dot\al +\dot\beta \right) \left[ \frac{\pa W}{\pa t} +\dot\al r\frac{\pa W}{\pa r} +\dot\al W\right] \notag \\
& \hspace{2.4in} =-2\left[\sigma +\frac{\rho_0}{U} \left( 2\dot\al +\dot\beta \right) r\right] \dot\al\Omega +\mu D_r\frac{\pa\Omega}{\pa r}
\end{align}
\bee
\left( 2\dot\al +\dot\beta \right)\left(\ddot\beta +\dot\beta^2 \right)=0.
\ene

Using equation (41), and the $\theta$-component of the equation of motion, namely
\bee
\rho\left(\frac{\pa W}{\pa t} +\dot\alpha r\frac{\pa W}{\pa r} +\dot\al W\right)=\mu\frac{\pa}{\pa r} (D_rW)
\end{equation}
equations (44) and (45) become
\bee
\left[\sigma -\frac{\rho_0}{U} \left(\frac{\dot\sigma}{\sigma}\right) r\right] \left[ \frac{\pa\Omega}{\pa t} +\dot\al r\frac{\pa\Omega}{\pa r} +2\dot\alpha\Omega\right]  - \frac{\frac{\rho_0}{U} \left( \frac{\dot\sigma}{\sigma} \right)}{\left[ \sigma - \frac{\rho_0}{U} \left( \frac{\dot\sigma}{\sigma} \right) r \right]} \mu \frac{\pa\Omega}{\pa r} =\mu D_r \frac{\pa\Omega}{\pa r}
\ene
\bee
\frac{\dot\sigma}{\sigma} \left(\ddot\beta +\dot\beta^2 \right)=0.
\ene

For the compressible case, $\dot\sigma / \sigma \neq 0$, so that equation (48) reduces to
\bee
\ddot\beta +\dot\beta^2 =0.
\end{equation}

In order to facilitate analytic calculation   consider again the case
\begin{gather}
\sigma (t)=\rho_0 e^{\int^t c(t')dt'} , \quad \dot\al (t)=-\frac12 a(t), \notag\\
 \dot\beta (t)=b(t).
\end{gather}

Equations (41), (43), and (49) then yield a closed set of equations for the   quantities $a(t)$, $b(t)$, and $c(t)$ (which are the same as those for the 2D case):
\begin{gather}
b(t)=\frac{1}{t+A}, \quad c(t)= \frac{e^{\frac12 \int^t a(t')dt'}}{B-\int^t e^{\frac12 \int^{t'} a(t^{\prime\prime} )dt^{\prime\prime}}dt'} , \notag\\ 
a(t)=b(t)+c(t) \quad
\end{gather}
where $A$ and $B$ are arbitrary constants.

Equation (47) then becomes
\bee
\left( \rho_0 e^{\int^t c(t')dt'} - \frac{\rho_0}{U} cr\right) \left( \frac{\pa\Omega}{\pa t} -\frac12 ar\frac{\pa\Omega}{\pa r}-a\Omega \right) 
 -\frac{(\frac{\rho_0}{U}) c}{\left( \rho_0 e^{\int^t c(t')dt'} 
-\frac{\rho_0}{U} cr \right) } \mu \frac{\pa\Omega}{\pa r} =\mu D_r \frac{\pa\Omega}{\pa r} .
\ene

In order to simplify equation (52), let us   assume again that the  compressibility effects are weak. From (41) and (50), this implies that the   quantity $c(t)$, which is a measure of the density change, is  small. Equation (43) then leads to
\bee
\frac{dc}{dt}-\frac12 a(t)c\approx 0
\end{equation}
from which
\bee
c(t)\approx c_0 e^{\frac12 \int^t a(t')dt'} 
\end{equation}
$c_0$ being an arbitrary constant. (54) replaces the second of the three solutions in (51). Further, in the weakly compressible case,   equation (48) (which, to first approximation, is   automatically satisfied) does not lead to   equation (49) which, therefore, has to be   abandoned. This implies that the first of the three solutions in (51), which comes from equation (49), then has to be dropped, as in the 2D case. 

Thus, keeping only terms of $O(c)$, equation (52) may be approximated by
\bee
\frac{\pa\Omega}{\pa t} - \frac{\nu c}{U} \frac{\pa\Omega}{\pa r} -\frac{ar}{2} \frac{\pa\Omega}{\pa r} \approx a\Omega + \nu \left( 1+\frac{cr}{U} \right) D_r \frac{\pa\Omega}{\pa r} 
\end{equation}
where
$$
\nu\equiv\frac{\mu}{\rho_0} .
$$

Let us look for a solution of the form (\'a la Lundgren \cite{tsl82})
\begin{gather}
\Omega (r,t)=S(t)\hat \Omega (\xi , \tau )\notag \\
\xi\equiv\sqrt{S(t)} \; r, \quad \tau\equiv \int_0^t S(t')dt', \quad S(t)=e^{\int_0^t a(t')dt'} .
\end{gather}

Equation (55) then becomes
\bee
\frac{\pa \hat \Omega}{\pa \tau} -\hat c \frac{\pa \hat \Omega}{\pa \xi} \approx  \nu \left( 1+ \frac{\hat c}{\nu} \xi \right) D_\xi \frac{\pa \hat \Omega}{\pa \xi}
\end{equation}
where
$$
\hat c (t)\equiv \frac{\nu c(t)}{U\sqrt{S(t)}} \quad D_\xi\equiv \frac{\pa}{\pa\xi} +\frac{1}{\xi} .
$$
The imposed straining flow has been transformed away by the  Lundgren transformation (56). Observe that the second term on the left hand side in   equation (57) represents the {\it first}\,-order   contribution due to the  compressibility effects   (assumed to be weak) -- the  compressibility effects impart a  hyperbolic character to   equation (57), as in the 2D case.

\medskip
\noindent 
{\bf (i)  Quasi-Steady Solution}

Equation (57) admits a quasi-steady   solution given by
\bee
\hat \Omega =\hat \Omega (\xi )
\end{equation}
which satisfies
\bee
\hat c \frac{\pa \hat \Omega}{\pa \xi} +\nu \left( 1+ \frac{\hat c}{\nu} \xi \right) D_\xi \frac{\pa \hat \Omega}{\pa\xi} \approx 0 .
\end{equation}

Equation (59) has the solution
\bee
\hat \Omega \approx CEi \left( \frac{\hat c}{\nu} \xi \right) =CEi \left( \frac{c(t)}{U} r \right)
\end{equation}
where $Ei(x)$ is the exponential integral
$$
Ei(x)\equiv\int_x^\infty \frac{e^{u}}{u} du,
$$
and $C$ is an arbitrary constant.

(60) has the following asymptotic behavior -
\bee
\hat \Omega \sim \frac{1}{r} e^{- \frac{c(t)}{U} r} , \quad r \;\; \text{large}.
\end{equation}

\noindent
The exponential decay of the vorticity for   large $r$ signifies the  enhanced compactness of the vortices  due to an {\it externally}-{\it imposed}  density build-up near the axis. One may in fact view (61) as a ``{\it dressed}\,"   vortex in analogy with the terminology in the  dielectric screening of a charged particle polarizing the surrounding medium (Ashcroft and Mermin \cite{am75})! ``{\it Dressed}\," vortex owes its extistence to a counter-conventional {\it externally}-{\it imposed} density build-up in the vortex core\footnote{Such vortices do not appear to be stable because the density increase in a direction opposite to that of the effective gravity due to the centrifugal force (which is directed away from the axis) would correspond to a top heavy arrangement under gravity. Indeed, swirling flows are found to be stabilized by a density stratification increasing in the radial direction (Howard \cite{lnh73}) while the vortex breakdown process is found to be delayed by the latter type of density stratification (Shivamoggi and Uberoi \cite{su81}).} (which is in contrast to a density drop in the vortex core in a normal compressible case).

\medskip
\noindent 
{\bf (ii) Unsteady Solution}

For the unsteady case, equation (57) has an approximate solution
\bee
\hat \Omega \approx f (\xi +\hat c \tau ) \frac{1}{\tau} e^{-\frac{\xi^2}{4\nu\tau}}
\end{equation}
where $f(x)$ is an arbitrary function of $x$. (62) may be viewed as a propagating   axisymmetric vortex in a compressible fluid.

If $a(t)=$ const $=a$,\footnote{The case $a(t)=$ const, according to (51), is valid only in the  weak-compressibility limit (small $c$), and  in the generic situation (arbitrary $c$) it is not valid.} using (62),(56) becomes  
\bee
\Omega\approx a \frac{f\left[ e^{\frac12 at} \{r+\frac{\nu c}{aU} (1-e^{-at} )\} \right]}{(1-e^{-at} )} \; e^{-\frac{ar^2}{4\nu (1-e^{-at})}} .
\end{equation}

Observe that (63), in the limit $t\Rightarrow\infty$,  gives the  axisymmetric steady Burgers vortex:
\bee
\Omega\approx\frac{\Gamma a}{4\pi\nu} e^{-\frac{ar^2}{4\nu}}
\end{equation}
where $\Gamma$ is the circulation around the vortex (and $f$ has been chosen suitably).

The azimuthal velocity corresponding to (64) is
\bee
W =\frac{\Gamma }{2\pi r} \left( 1-e^{-\frac{ar^2}{4\nu}} \right) .
\end{equation}
(65) describes a rigid-body rotation for small $r$, and an  irrotational flow field for large $r$. The azimuthal velocity is maximum for 
$r=r_* \sim \sqrt{\nu /a}$. Thus, $r_*$, which may be taken to be radius of the vortex core, is of the order of Kolmogorov microscale $\eta \sim (\nu^3 /\epsilon )^{1/4}$, if $a\sim \sqrt{\epsilon / \nu} \, ,$ $\epsilon$ being the energy dissipation rate in turbulence.

\section{Applications to Turbulence}

{\bf (i) Incompressible Turbulence}

(64) implies that the relevant length scale for the incompressible case is
\bee
\ell^2 \sim \frac {\nu}{a}.
\ene

Taking the core radii of Burgers vortices to be of the order of  Kolmogorov microscale we have
\bee
a\sim\epsilon^{1/2} \nu^{-1/2}
\ene
where,
\bee
\epsilon\sim\nu\frac{U^2}{\ell^2}.
\ene
(66)-(68) lead to 
\bee
U\sim\epsilon^{1/3}\ell^{1/3}
\ene
and hence to the celebrated Kolmogorov \cite{ank41} spectrum for incompressible turbulence 
\bee
E(k)\sim\epsilon^{2/3}k^{-5/3}.
\ene 
 
\noindent 
{\bf (ii) Compressible Turbulence}  

(37) and (63) imply that the relevant length scale for the  compressible case is 
\bee
\ell\sim\frac{\nu c}{aU} .
\ene

Recalling that $c\Rightarrow0$ corresponds to the   incompressible limit, we may write
\bee
c\sim\frac{MU}{\ell}
\ene
where $M$ is a reference Mach number of the flow
\bee
M\sim\frac{U}{\ C}
\ene
$C$ being a reference speed of sound.

Further, writing
\bee
a\sim\frac{U}{\ell}
\ene
we have, from (71),
\bee
\ell\sim\frac{\nu}{C} .
\ene

On noting now that the energy dissipation rate can be written as
\bee
\hat\epsilon\sim\mu \frac{U^2}{\ell^2}
\ene
we obtain, from (75),
\bee
U\sim \rho^{-1/2} \hat\epsilon^{1/2} C^{-1/2} \ell^{1/2}
\ene
which leads to the Kadomtsev-Petviashvili \cite{kp73} spectrum for compressible turbulence
\bee
E(k)\sim \hat \epsilon C^{-1} k^{-2}.
\ene

\section{ Vortex Reconnection in a Compressible Fluid}

We consider a generalization of Greene's \cite{jmg93} {\it local vortex pseudo-advection} (the   terminology is, however, due to Vallis et al. \cite{vcy90}) to make a general discussion of the vortex reconnection process in a compressible fluid.
 
The vorticity evolution equation in an inviscid fluid is
\bee
\frac{\pa {\boldsymbol \omega}}{\pa t}+\nabla \times ({\boldsymbol \omega} \times {\bf v} )=-\nabla \times \left(\frac{1}{\rho} \nabla p\right).
\ene
The term on the right represents {\it baroclinic} vorticity generation which is due to the   misalignment of density and pressure gradients. Note that for an incompressible or a   compressible barotropic fluid this term vanishes,   so the vorticity evolution in an  inviscid   incompressible or a compressible barotropic fluid is simply a {\it local vortex advection} signifying the absence  of vortex reconnection.

On the other hand, for a compressible non-barotropic fluid, if
\bee
{\boldsymbol \omega} \cdot \frac{1}{\rho} \nabla p=0, \quad \forall \, \bf x\in {\mathscr V}
\ene 
$\mathscr V$ being the volume occupied by the fluid, i.e., the  vortex lines are confined to the isobaric \linebreak ($p=$ const) surfaces, then one may write
\bee
\frac{1}{\rho} \nabla p={\boldsymbol \omega } \times {\bf W} , \quad\forall \, \bf x\in {\mathscr V} .
\ene
Equation (79) then becomes
\bee
\frac{\pa {\boldsymbol \omega}}{\pa t} +\nabla\times [{\boldsymbol \omega} \times ({\bf v} + {\bf W})] =0.
\ene
(82) implies that, under condition (80), the vorticity evolution in a compressible non-barotropic fluid  corresponds to a {\it local vortex   pseudo-advection} by a modified velocity ${\bf v}+{\bf W}$ where,  from (81),
\bee
{\bf W} =\frac{1}{\rho \, \omega^2} \nabla p\times{\boldsymbol \omega} .
\ene

Further, the helicity
\bee
H\equiv {\boldsymbol \omega} \cdot {\bf v}
\ene
which is a topological measure of the degree of knottedness of vortex lines, then evolves according to
\bee
\frac{\pa H}{\pa t} +\nabla \cdot [({\bf v} +{\bf W} )H] =\nabla\cdot \left[ {\boldsymbol \omega} \left( H+\frac12 v^2 \right)\right] .
\ene

Integrating equation (85) over the volume ${\mathscr V} (t)$ enclosed by a surface $S(t)$ moving with   velocity ${\bf v} +{\bf W}$ on which ${\boldsymbol \omega} \cdot \hat {\bf n} =0$ (i.e., $S(t)$ is a vortex surface, as implied by equation (82)), we obtain
\bee
\frac{d}{dt} \int\limits_{\mathscr V(t)} Hd{\bf x} =0.
\ene
So, provided (80) is valid, the total helicity is conserved, even in a compressible non-barotropic fluid, despite the existence of  {\it baroclinic} vorticity generation mechanism.

It should be noted however that the prevalence of   {\it local vortex pseudo-advection} and hence the absence of vortex reconnection is a {\it sufficient} (but {\it not} necessary) condition for conserving the  total helicity also in a compressible fluid. Therefore, the absence of {\it local vortex pseudo-advection} and hence the occurrence of vortex reconnection does not guarantee the destruction of the total helicity invariant.

In the generic compressible non-barotropic case, where (80) is not valid, the vorticity evolution does not correspond to a {\it local vortex pseudo-advection}. This paves the way for the occurrence of   vortex reconnection in a compressible non-barotropic fluid even in the {\it inviscid} case! DNS of the reconnection process between two anti-parallel vortex tubes (Virk et al. \cite{vhk95}) in fact showed that {\it shocklet} formation was able to get reconnection going in a compressible fluid.

{\it Inviscid} \,compressible vortex reconnection is very akin to the  {\it collisionless}   magnetic reconnection process in high-temperature tenuous plasmas where resistivity is negligible (Coppi \cite{bc64}, Schindler \cite{ks74}, Drake and Lee \cite{dl77}, Ottaviani and Porcelli \cite{op95}, Shivamoggi \cite{bks97}-\cite{as03}). Here, the conservation  of magnetic flux is   replaced by the conservation of {\it generalized}  magnetic flux (that now includes contributions from the electron-fluid momentum). So, magnetic flux changes and magnetic reconnection processes are sustainable even without resistivity!

\section{Discussion}

In this paper, stretched (modified Burgers) vortices are considered in a compressible fluid. The flows associated with the vortices are    {\it perpendicular} to the plane of the uniform straining flows -- a situation relevant for a mixing-layer flows. Compressibility effects have been restricted to be weak to facilitate analytic solutions. The  compressible axisymmetric stretched   vortex - 
\begin{itemize}
\item exhibits exponential decay of the vorticity for large $r$ signifying the enhanced compactness of the vortices caused by an {\it externally-imposed}  density build-up near the axis -- ``dressed" vortices,
\item has the axisymmetric Burgers vortex as the asymptotic limit $(t\Rightarrow\infty)$.
\end{itemize}

The compressible vortex flow solutions in the 2D as well as axisymmetric cases  identify  a length scale relevant for the   compressible case which  leads to the    Kadomtsev-Petviashvili \cite{kp73}  spectrum   for  compressible turbulence.  

Vortex reconnection in  a compressible non-barotropic fluid  is possible even in the {\it inviscid}   case -- compressibility leads to defreezing of vortex lines in the fluid. This is very similar to the {\it collisionless}   magnetic reconnection process in  high-temperature  tenuous plasmas.

The possibility of vortex reconnection in an {\it inviscid} fluid can raise some questions of principle (\'a la Taylor, as quoted in \cite{op95}, for the {\it collisionless} magnetic reconnection process). Since the process is {\it reversible} one might wonder if the reconnection in such a system is only a transient phenomenon and if the vortex lines will eventually unreconnect. However, the essential presence of even a very small viscosity would inhibit the latter process.

\bigskip

\noindent
{\large {\bf Acknowledgments}}

My thanks are due to Peter Constantin, Klaus Els\"asser,  Bob Kerr,  Keith Moffatt, and Mahinder Uberoi for valuable remarks and suggestions.

\end{document}